\documentclass[10pt,journal,onecolumn]{IEEEtran}
\ifCLASSINFOpdf
 
\else
 
\fi

\usepackage{amsmath,amsfonts}
\usepackage{algorithmic}
\usepackage{graphicx}
\usepackage{textcomp}
\usepackage{pifont}
\usepackage{tikz}
\usepackage{caption}
\usepackage{subfig}
\usepackage{gensymb}
\usepackage{url}
\usepackage{amsmath}
\usepackage{amsfonts}
\usepackage{float}
\usepackage{graphicx}
\usepackage{multirow}
\usepackage{mathtools}
\usepackage{enumitem}
\setlist{leftmargin=*}
\usepackage{colortbl}

\makeatletter
\renewcommand\footnoterule{%
  \kern-3\p@
  \hrule\@width 0.5\columnwidth
  \kern2.6\p@}
  \makeatother
\begin{document}

\title{DeMiST: \underline{De}tection and \underline{Mi}tigation of \underline{S}tealthy Analog Hardware \underline{T}rojans}

\author{
    \IEEEauthorblockN{
        Enahoro Oriero\IEEEauthorrefmark{3}, Faiq Khalid\IEEEauthorrefmark{2}\textsuperscript{*}\thanks{\textsuperscript{*}Faiq Khalid and Enahoro Oriero equally contributed to this paper}, Syed Rafay Hasan\IEEEauthorrefmark{3}\\ 
    }
    \IEEEauthorblockA{
        \IEEEauthorrefmark{3}\textit{Tennessee Tech. University, Cookeville, TN, USA}\\
        \IEEEauthorrefmark{2}\textit{Technische Universit\"at Wien (TU Wien), Vienna, Austria}\\    
        eoriero42@students.tntech.edu, faiq.khalid@tuwien.ac.at, shasan@tntech.edu, 
    }
}

\markboth{Accepted at ACM Hardware and Architectural Support for Security and Privacy (HASP) 2023}{E. Oriero \MakeLowercase{\textit{et al.}}: DeMiST}

\maketitle

\begin{abstract}
The global semiconductor supply chain involves design and fabrication at various locations, which leads to multiple security vulnerabilities, e.g., Hardware Trojan (HT) insertion. Although most HTs target digital circuits, HTs can be inserted in analog circuits. Therefore, several techniques have been developed for HT insertions in analog circuits. Capacitance-based Analog Hardware Trojan (AHT) is one of the stealthiest HT that can bypass most existing HT detection techniques because it uses negligible charge accumulation in the capacitor to generate stealthy triggers. 

To address the charge sharing and accumulation issues, in this paper, we first propose a novel way to detect such capacitance-based AHT. Secondly, we critically analyzed existing AHTs to highlight their respective limitations. By addressing these limitations, we proposed a stealthier capacitor-based AHT (fortified AHT) that can bypass our novel AHT detection technique. Finally, by critically analyzing the proposed fortified AHT and existing AHTs, we developed a robust two-phase framework (DeMiST) in which a synchronous system can mitigate the effects of capacitance-based stealthy AHTs by disabling the triggering capability of AHT. In the first phase, we demonstrate how the synchronous system can avoid the AHT during run-time by controlling the supply voltage of the intermediate combinational circuits. In the second phase, we proposed a supply voltage duty cycle-based validation technique to detect capacitance-based AHTs. Furthermore, DeMiST amplified the switching activity for charge accumulation to such a degree that it can be easily detectable using existing switching activity-based HT detection techniques. 

To illustrate the effectiveness, we evaluated the proposed fortified AHT and two-phase AHT detection on state-of-the-art benchmark circuits. The experimental results show that with less than 4\% area overhead, the proposed detection technique successfully detects capacitance-based AHTs.

\end{abstract}


%=================================================================================
\section{Introduction}\label{sec1}
The competitive nature of Integrated Circuits (IC) design makes the time-to-market and affordability of ICs very significant factors in the IC design cycle. In addition, variations in the performance requirements and ever-evolving end-user applications increase the complexity and production cost of IC development. These challenges force IC design companies or system-on-chip (SoC) integrators to outsource some of the steps in the design process to untrusted third-party intellectual property (3P-IP) vendors or use untrusted commercial-off-the-shelf components. Hence, it makes SoCs prone to hardware Trojans (HT) insertion from 3P-IP, which can be supplied to SoC integrators as hard IP, soft IP, or firm IP and can use digital or analog triggers. HTs use external/internal signals or logic. Alternatively, HTs can be in an always-on state to degrade the performance or to trigger the malicious payload (functional manipulation, information leakage, and denial of service)~\cite{a5,a4}. The outcome of these HTs can lead to catastrophic incidents and financial loss. \textit{Hence, it is imperative to study the vulnerabilities of ICs against all sorts of analog or digital HTs}.

Towards the vulnerabilities analysis, researchers have proposed several HTs in the literature that exploit different vulnerabilities in ICs to add additional circuitry or exploit the existing circuits to change the properties (power, temperature, and delay) and functionality. Furthermore, HTs control existing circuitry or insert a stealthy circuit to develop a trigger~\cite{n1,n2,n3,n4,n5,n6,n7} that activates different payloads to perform the targeted malicious activities. Most HT insertion approaches generally come with additional area and power overhead. Although some HTs exhibit almost negligible area and power overhead (like TrojanZero~\cite{n4}), most of them consist of digital payloads and triggers. In literature, an alternate approach is to use analog circuits as triggers to reduce the area and power overhead. These kinds of HTs are based on the charge accumulation of a capacitor (only one transistor is used as a capacitor) over many clock cycles. The charge accumulation is achieved by exploiting the switching activity over many clock cycles. The accumulated charge acts as a trigger signal to the associated analog or digital payload. In this paper, we called charge accumulation-based HTs Analog Hardware Trojan (AHT). The charge accumulation-based triggers reduce the area and power overhead in an HT and also require minimal perturbations in the original circuit~\cite{a8}. In addition, AHT redirects and accumulates small amounts of charge from the victim wire (i.e., the output node of a logic element) and feeds the trigger circuit's capacitor. Since AHT requires no extra logic, and the basic idea requires the accumulation of charges, traditional HT detection techniques, like increasing the probability of rare transition occurrence or side-channel analysis, cannot detect AHTs~\cite{ab}. Furthermore, the AHT camouflages its operation if successfully deployed within a combinational design. The reason behind camouflaging the HT operation is that charge accumulation in the combinational circuit due to a wire's signal transition is less controllable than a sequential circuit element's node, which usually reacts at every clock edge. All these characteristics of AHTs and limitations of the existing HT detection technique lead to the following research question: 

\textit{\textbf{RQ1:} How can we stop charge accumulation without affecting normal operations?}

To address this research question, we propose a novel charge-depriving technique to detect AHT based on charge accumulation in a capacitor. This technique depletes the VDD to starve the capacitor from charging while ensuring normal operation. To evaluate the effectiveness of the techniques, in this paper, we utilize the state-of-the-art $A2$ AHT (reported in \cite{a8}) as a benchmark and evaluate our proposed AHT detection. The experimental analysis shows that charge-depriving techniques are only applicable if capacitor-based HT is directly controlled through supply voltage. This leads to the following research question:

\textit{\textbf{RQ2:} Can the charge-depriving against the defense be exploited to affect normal operation?} 

By exploiting the limitation of the proposed charge-depriving technique, we critically analyzed our proposed AHT detection technique. In the process, we designed a fortified AHT that is more stealthy and robust and still operates under charge-depriving conditions. This robust AHT exploits the concept of charge sharing and introduces another VDD-controlled pMOS transistor to share the charge accumulation. Although it increases the area overhead by one transistor, it reduces the charge accumulation requirement for an additional transistor(s) (which acts as a capacitor). This reduction allows the attacker to combine the total accumulated to obtain the required accumulated charge that can be strong enough to generate the triggering signal. Since this technique requires minimal charge accumulation, it can still operate under the charge-depriving phase of the proposed AHT detection. This leads to the following research question:

\textit{\textbf{RQ3:} How do we detect and mitigate capacitor-based AHTs that exploit the charge-depriving technique with negligible overhead?}
 
To address RQ3, we proposed a robust two-phase framework AHT detection solution (DeMiST) where any AHT may co-exist with a synchronous system while performing standard functionality. However, DeMiST amplified the switching activity for charge accumulation to such degree that it can be easily detectable using existing switching activity-based HT detection techniques like layout-aware switching activity localization~\cite{a27} and side-channel activity magnifier~\cite{a28}. In addition, DeMiST mitigates the AHT by disabling the AHT capability to generate the trigger or activate its malicious payload. In summary, DeMisT is composed of two phases: design-time mitigation and Pre-market validation. The first phase involves starving any possible capacitor responsible for the AHT from charging, preventing them from triggering. The second phase of our technique provides a complete framework that can detect AHT at the pre-market test stage and mitigate the effect of AHT. \textbf{In summary, the novel contributions of this paper are:}

\begin{enumerate}
    \item \textbf{Chrage-depriving-based AHT detection (see Section~\ref{sec5}):} To address RQ1, we propose a novel charge-depriving technique to detect capacitor-based AHT that exploits the negligible charge accumulation for triggering.
    \item \textbf{Fortified AHT (see Section~\ref{sec6}):} To address RQ2, we proposed a robust AHT that bypasses the charge-depriving technique by adding a VDD-powered pMOS using the charge-sharing technique, which further reduces the required charge accumulation that requires for triggering.
    \item \textbf{DeMiST (see Section~\ref{sec7}):} To address RQ3, we proposed a two-phase framework to detect and mitigate the effects of the fortified AHT in a post-fabricated IC. This framework allows the AHT to co-exist in a traditional synchronous system at the cost of a nominal performance penalty. 
\end{enumerate}

To show the effectiveness, we evaluated the proposed capacitance-based AHT and two-phase AHT detection on state-of-the-art benchmark circuits, i.e., s298, s344, c432, and c880. The experimental results show that with less than 4\% area overhead, the proposed detection technique successfully detects capacitance-based AHTs. For example, the area overhead of our proposed $VDD$ control circuit is as little as $1.65$\% if introduced in a moderately complex digital logic such as ISCAS benchmark circuit c880.
\section{Related Work}\label{sec2}
Hardware Trojan attacks have emerged as a major security concern for integrated circuits (ICs), and many recent papers have provided comprehensive overview \cite{a9,ac,ad}. As stated earlier, HT typically consists of a payload delivering the malicious functionality and an activating triggering circuit.

Much of the existing literature focuses on HT detection based on diverse characteristics such as parametric (physical) properties, trigger, and payload features \cite{a10}.
One widely researched technique is using side-channel analysis to detect abnormal behavior due to HT.
This is based on the premise that a malicious addition during design or fabrication is bound to have an effect on the power consumption profile or timing profile of the circuit  \cite{a11}. 
The detection sensitivity of side-channel test methods depends upon the precise measurement of side-channel signals.

To overcome this limitation, logic testing approaches have been introduced. In logical testing, a test method is derived to identify the Trojan circuit by monitoring the output responses when effective functional test patterns are exerted to stimulate the design circuit. The goal is to detect low-probability conditions at the internal nodes and then derive an optimal set of vectors that can trigger each of the selected low-probability nodes individually to their rare logic values \cite{a16}. 
The HT is detected if it affects the IC’s response during the test procedure.
Siraj et al. in \cite{a17a,a17b}, proposed a self-triggering HT where an intruder can leverage the exacerbated negative bias temperature instability effect to trigger the Trojan payload. So far, lesser emphasis is made on analog circuit-based hardware Trojan.
Another set of AHT detection techniques is laser logic state imaging. Although this technique works effectively, it exhibits a high computational and resource cost~\cite{n8,n10}. 
In \cite{a8,n9}, traditional parameterizing-based techniques demonstrate that AHT is difficult to detect. 
This is due to the minuscule level of changes required to introduce AHT, which usually falls within the noise budget of conventional parametric sensing. To the best of the authors’ knowledge, no viable solution is provided to detect AHTs. Therefore, in this research, we undertake this problem.

\section{Threat Model}\label{sec3}
This paper considers a threat model based on the following factors: (1) Which part of the SoC can be affected? (2) Who are the attacker and defender, and what are their access capabilities? 

\begin{itemize}
    \item We considered a standard threat model used in literature, in which IPs are developed at untrusted 3P-IP vendors, and the attacker can be at the third-party site inserting malicious HT into the IPs.

    \item During the system-on-chip (SoC) integration, malicious IPs can be unknowingly integrated into the system design, affecting functionality that leads to catastrophic incidents or financial loss.

    \item The SoC integration facility is considered trusted, and all the HT mitigation or detection techniques can only be added at the SoC integration stage. Therefore, during design, the IP designer includes an extra port after each pipeline stage of the design.

    \item We assume that the attacker has no knowledge or control over the integration and post-fabrication procedures.

    \item Post-fabrication pre-market testing is done in a trusted facility.
\end{itemize}

It is important to note that all the above-mentioned assumptions are taken from the real-world SoC supply chain model~\cite{ae}, which is most-commonly used in the literature.
\section{Analog Hardware Trojan (AHT): A Review and its Robustness}\label{sec4}
\begin{figure}[!ht]
    \centering
    \includegraphics[width=0.5\linewidth]{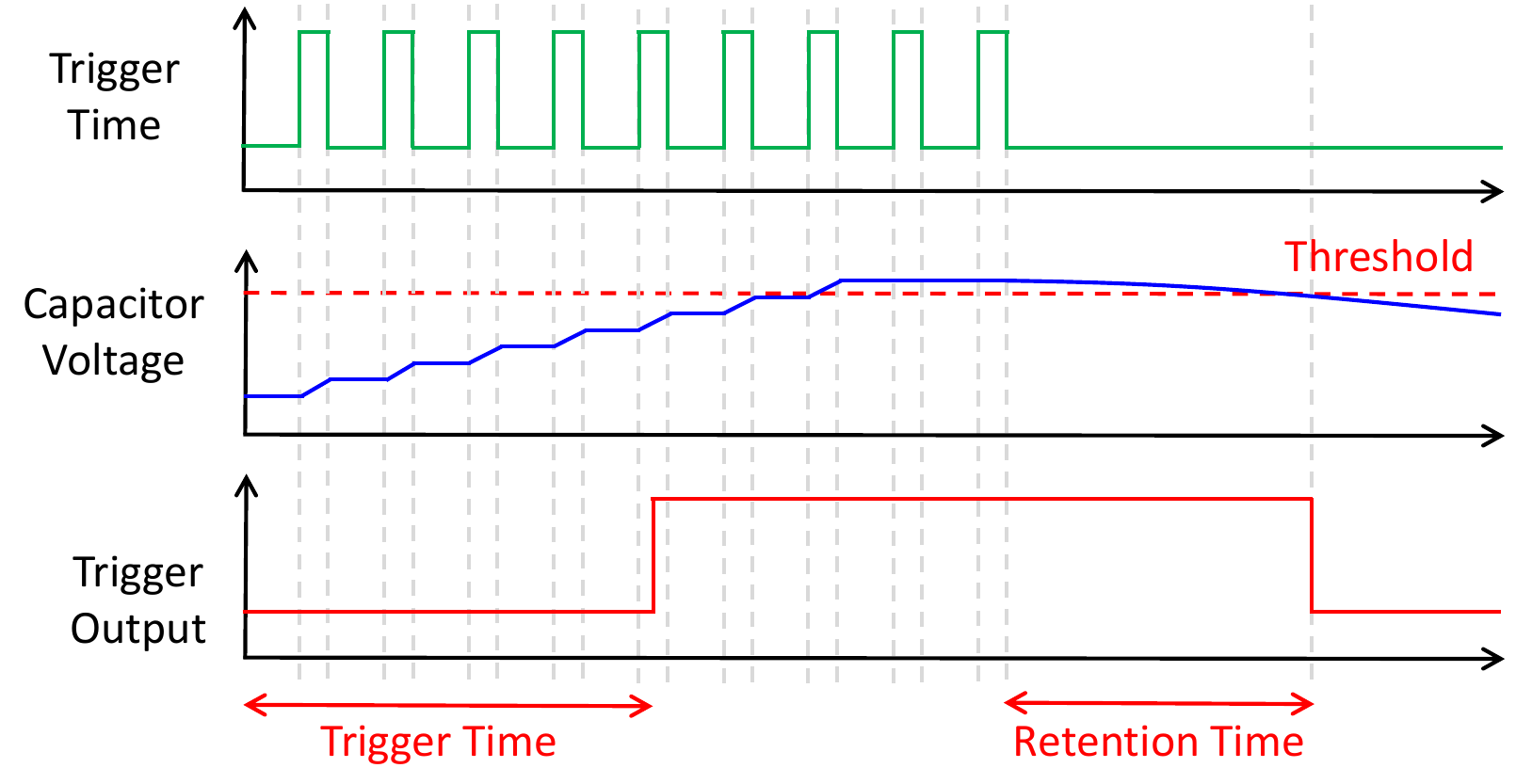}
    \caption{Traditional analog trigger model that charges the capacitor to a certain threshold, which acts as a trigger signal for the Trojan payload~\cite{a8}.}
    \label{fig:ATM}
\end{figure}

To the best of our knowledge, Yang et al. proposed an AHT model, A2, for the first time \cite{a8}. 
In this section, we examine the $A2$ AHT. 
Fig. \ref{fig:ATM}, shows the behavioral model of the analog trigger circuit.
The basic idea behind this model is based on charge accumulation in a capacitor due to frequent transistor switching. 
Each time the victim wire that feeds the trigger circuit’s capacitor toggles (i.e., changes value), the capacitor voltage changes. This change is represented as $\Delta$$V$. 
After a number of toggles, the capacitor’s voltage exceeds a predefined threshold voltage and enables the trigger output \cite{a8}.

The trigger time\footnote{It can be calculated by multiplying the toggling frequency of the input victim wire by the number of consecutive toggles to fill the capacitor.} is defined as the time it takes to activate the trigger fully.
When the trigger input is inactive, the capacitor’s voltage depreciates gradually due to its natural leakage property.
The retention time in Fig.~\ref{fig:ATM}(a) denotes the time it takes to reset the trigger output after the triggering input stops toggling. The retention time is therefore dependent upon the leakage current~\cite{a8}. The leakage current slowly reduces the accumulated charge and eventually reduces its value to lower than the threshold value, as shown in Fig.~\ref{fig:ATM}.

\begin{figure}[!ht]
    \centering
    \includegraphics[width=0.5\linewidth]{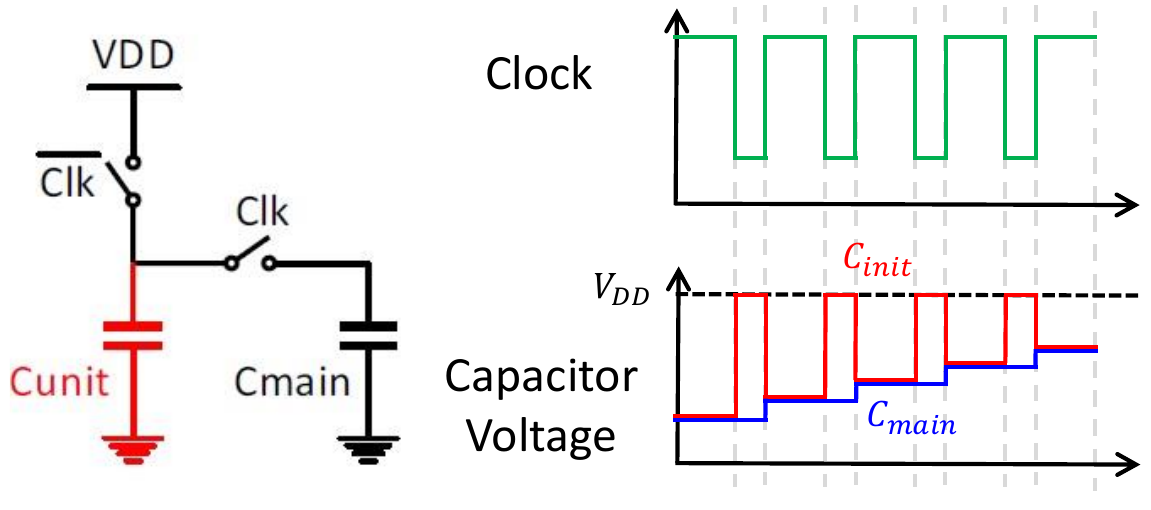}
    \caption{Analog trigger circuit based on capacitor charge sharing~\cite{a8}}
    \label{fig:ATC}
\end{figure}

\begin{figure}[!ht]
    \centering
    \includegraphics[width=0.7\linewidth]{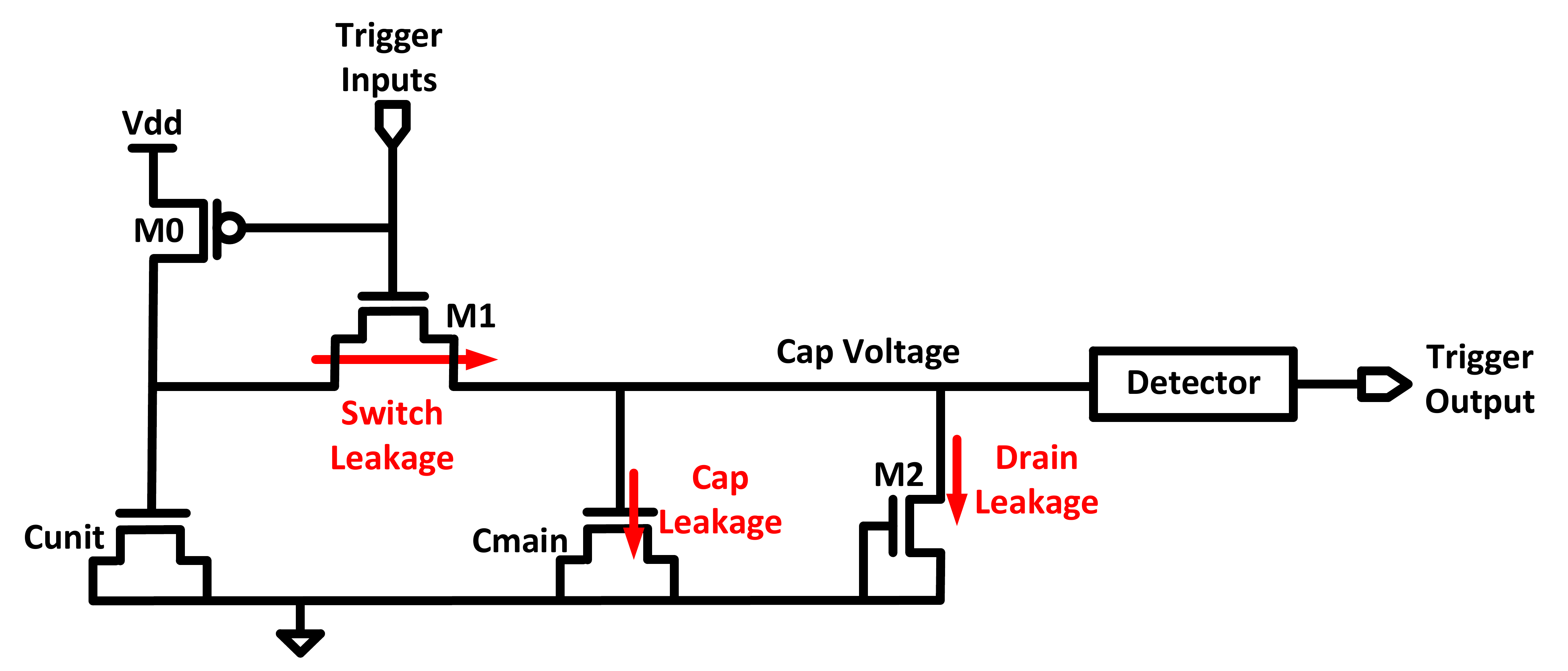}
    \caption{An example of analog hardware Torjan: Triggering input with detector circuit to assert the trigger output when $C_{main}$ stores enough charge~\cite{a8}.}
    \label{fig:AH_example}
\end{figure}

The physical implementation of the analog trigger model is shown in Fig. \ref{fig:ATC}.
Capacitor $Cunit$ controls the amount of charge dumped on the main capacitor $Cmain$.
When clock ($Clk$) remains low, $Cunit$ gets charged to $VDD$.  
When the $CLK$ signal turns high, the two capacitors ($C_{unit}$, and $C_{main}$) make a parallel circuit and share the available charges. Thus, the final voltage of the two capacitors is the same. The voltage change, $\Delta$$V$, on $Cmain$ is calculated using equation \ref{eq1}.
  
\begin{equation} \label{eq1}
\Delta V=\frac{Cunit \times (VDD - V_{0})}{Cunit + Cmain}
\end{equation}

where $V_{0}$ represents the initial voltage on $C_{main}$ before the switching of the transistor.
Yang et al. implemented this concept using the transistor-level circuit shown in Fig. \ref{fig:AH_example}. 
The two switches for $\overline{Clk}$ and $Clk$ in Fig. \ref{fig:ATC} are replaced by $M0$ and $M1$ in Fig. \ref{fig:AH_example}, respectively. 
To balance the leakage current through transistors $M0$ and $M1$, an additional leakage path $M2$ to the ground is added. 
The detector circuit senses whether the capacitor voltage has reached the threshold level and triggers accordingly. 

\section{Novel Testing Scheme to detect AHT}\label{sec5}
Due to the requirement of capacitor charging, traditional test vectors fail to trigger these AHT \cite{a21}. 
Similarly, due to a very small footprint, AHT cannot be detected using time fingerprinting or power profiling techniques \cite{a22}.
Therefore, this paper proposes a novel technique to detect AHT through a forced charge-depriving technique.

As can be seen in equation \ref{eq1}, the capacitor charging depends on the voltage source ($VDD$), and by depleting the $VDD$ supply, we can starve the capacitors from getting charged.
Depletion of $VDD$ supply prevents the capacitors from charging and correspondingly avoid triggering.

\begin{figure}[!ht]
    \centering
    \includegraphics[width=0.9\linewidth]{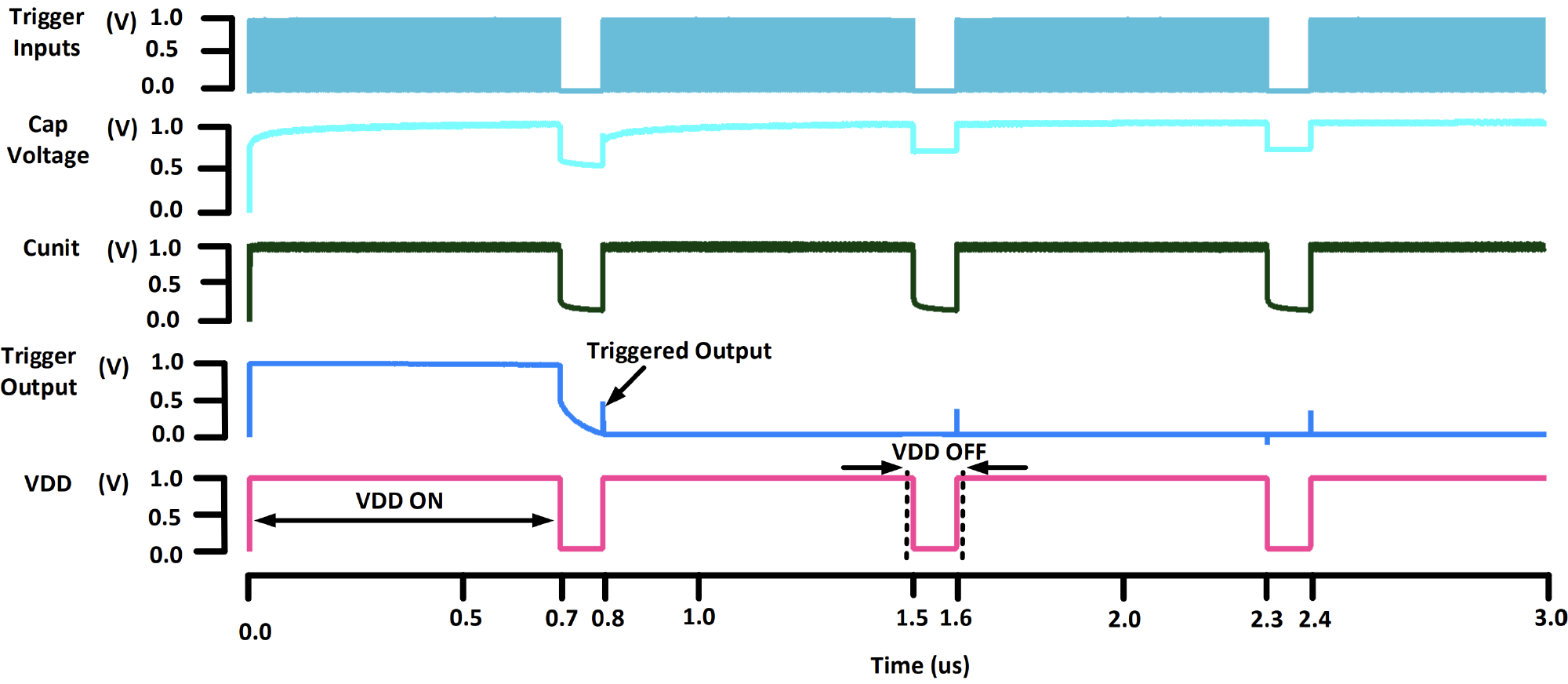}
    \caption{Timing waveform for AHT in Fig.~\ref{fig:AH_example} for test case 1.}
    \label{fig:ATW1}
\end{figure}

\textbf{Test Case 1:}
Fig. \ref{fig:ATW1} illustrates a test scenario where the supply voltage ($VDD$) is cut off before the trigger time which, the value of this trigger time is $770ns$ in our case, further details of simulation (including transistor sizing) are provided later on in this section.
%at $700ns$ (before the trigger time of $770ns$, which we obtained beforehand for our circuit)) and restored back at $800ns$.
The $VDD$ remains cut off for $100 ns$; again, this cut-off duration is dependent on the implementation. 
We discovered that for the given capacitances in our circuit, it takes about $100 ns$ for the circuit to restore the charges to a level that keeps the next triggering duration close to the same value, which is $700 ns$ in our case.
Therefore, $VDD$  is restored back to $800ns$. The depletion and restoration of $VDD$ at $700ns$ and $800ns$ correspondingly affected the trigger input, causing it to be active only when $VDD$ is available.

We implemented and simulated it using the $45$-$nm$ Predictive Technology Model on Cadence EDA Tool Suite.
For the simulation,  $VDD$ of $1V$ and temperature of 27\degree$C$ is kept.
%Typical Typical corner case is used for the transistors,
For the sizing, transistor $M0$ is $2\lambda$ and $M2$ is $4\lambda$.
$M1$, $Cunit$ and $Cmain$ are all $1\lambda$.
The trigger time is $770ns$ and the retention time is $30$$\mu$$s$.
The reason for the long retention time is because of the inability of $M2$, $Cunit$ and $Cmain$ to discharge quickly. 
This can be modified to shorten the retention time by decreasing the size of $Cunit$ and $Cmain$ transistors and increasing the size of the drain leakage transistor $M2$. 

As earlier explained, when transistor $M0$ is turned off, $Cunit$ and $Cmain$ share the available charges.
Similarly, when active, it constantly provides a path for $VDD$, which continuously dumps charges on the output capacitor represented by $Cap$ $Voltage$ in Fig. \ref{fig:AH_example}.
As shown in the waveform of Fig. \ref{fig:ATW1}, the output does not trigger during the first $700ns$ because the voltage of the capacitor does not charge up to the predefined threshold voltage.
On restoring $VDD$ at $800ns$, the output triggered instantly, indicating the presence of the Trojan.
This is indicated by the black arrow in Fig. \ref{fig:ATW1} (Triggered output).
This occurred because the gradual voltage increment caused the voltage at the output capacitor to cross the predefined threshold, which enabled the detector circuit to trigger the output.
Hence, we can conclude that when the trigger input is inactive, only transistor $M0$ is operating, causing unequal charge sharing between $Cunit$ and $Cmain$ and when active, it continuously charges $Cunit$. This unbalance charge distribution can be used to raise a detection flag. 

\begin{figure}[!ht]
    \centering
    \includegraphics[width=0.85\linewidth]{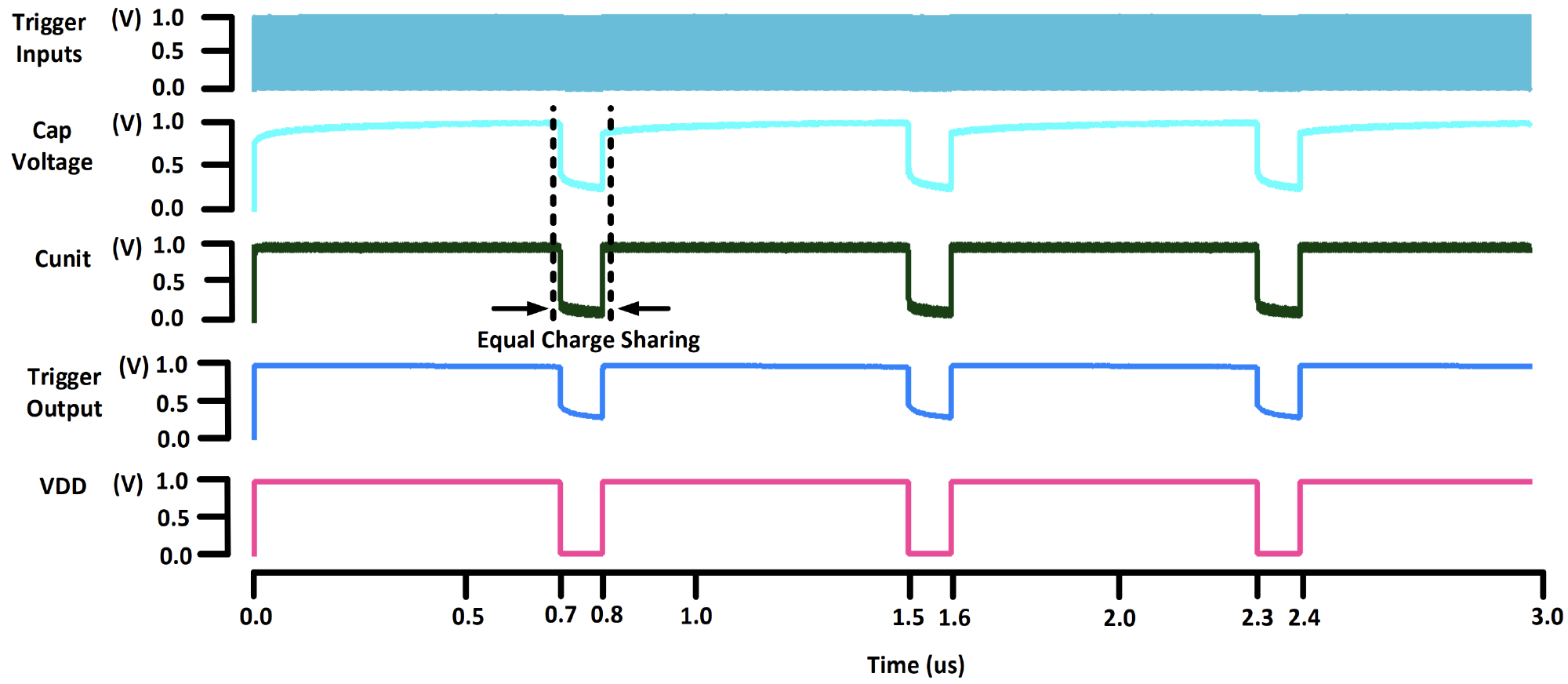}
    \caption{Timing waveform for AHT in Fig.~\ref{fig:AH_example} for test case 2.}
    \label{fig:ATW2}
\end{figure}

\textbf{Test Case 2:}
Fig. \ref{fig:ATW2} illustrates a second test case where the supply voltage ($VDD$) is similarly cut off at $700ns$ and restored at $800ns$ but a continuous Trigger input is considered. 
This is to undertake the case in which somehow the hidden triggering unit is not controllable by the regular VDD. 
The feasibility of this sort of scenario is further explained in Section 7.
From the waveform of Fig. \ref{fig:ATW2}, we notice that the output does not trigger (i.e., the Trigger output signal does not reach $0$ volt when $VDD$ is $0$ volt) and it responds according to $VDD$.
This occurs because the constant triggering input ensures there is always equal charge sharing between $Cmain$ and $Cunit$.
This also made sure that the voltage at the output capacitor remained within the threshold boundaries.
It illustrates that if the trigger input is continuous, judicious usage of the depletion of the supply voltage can prevent the  AHT from triggering the payload.

These results show that Test Case $1$ is only valid if the trigger input can be directly controlled through $VDD$. 
Test Case $2$, on the other hand, leads to camouflaging the AHT. 
However, our proposed circuit configuration is preventing such AHT from triggering. 
Therefore, as long as we calibrate the $VDD$ duty cycle correctly  (further explained in Section 7), the synchronous system can work normally even in the presence of the AHT.  
In the next section, we are proposing a more robust  AHT. Afterward, we'll discuss the mitigation of such robust AHT and how to calibrate $VDD$.   

\section{AHT Fortified}\label{sec6}
\begin{figure}[!ht]
	\centering
	\subfloat[Our fortified AHT with controlled $VDD$ and equal charge distribution]{{\includegraphics[width=0.7\linewidth]{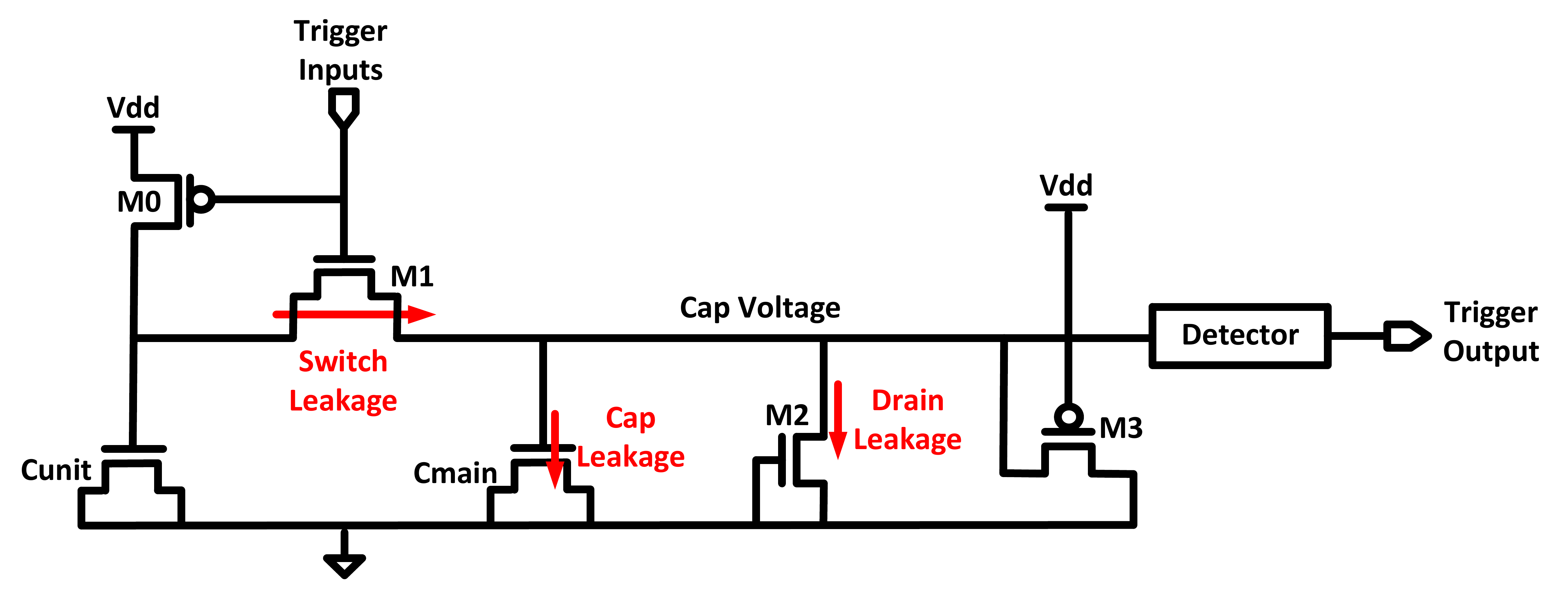} }}
	\qquad
	\subfloat[Analog trigger circuit with $C_{new}$]{{\includegraphics[width=0.7\linewidth]{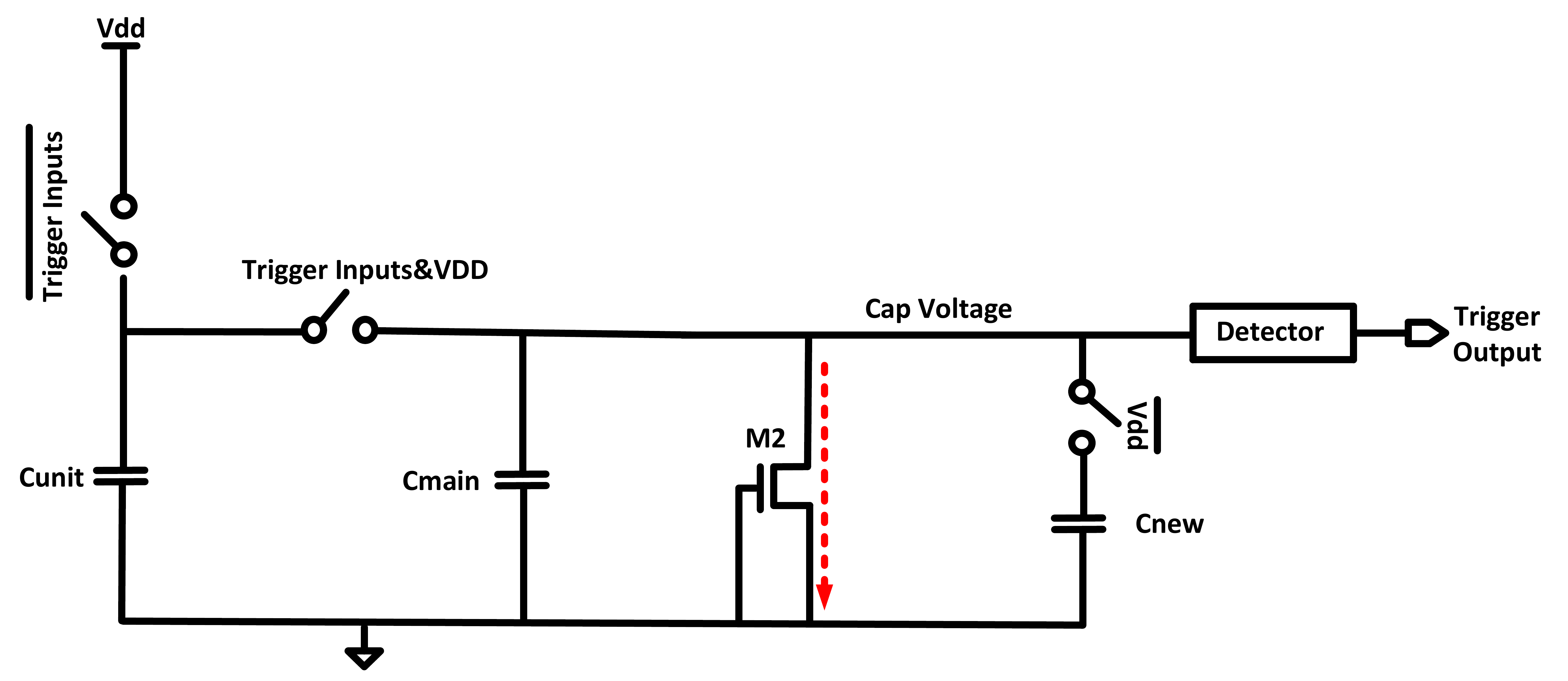} }}
	\caption{Transistor schematic of fortified analog trigger circuit and its RC model}
	\label{fig3}
\end{figure}

In this section, we are playing devil's advocate and proposing a fortified version of AHT. 
Fig. \ref{fig3}a shows the transistor schematic of the fortified AHT, which is enhanced by adding an extra $VDD$ controlled transistor ($M3$).
The extra PMOS transistor defeats the HT detecting methodology proposed earlier in this paper by ensuring equal charge sharing at all times.
Fig. \ref{fig3}b shows the RC model of this fortified AHT.

When the triggering input is low, $Cunit$ is charged to $VDD$.
$Cmain$ is charged when the Trigger input is turned on, and $VDD$ is asserted. 
During the time when $VDD$ is off, the $Cmain$ is discharged through additional parallel capacitance $Cnew$, which shares the charge with $Cmain$. 
To avoid any confusion,  we want to emphasize that  $Cnew$ represents capacitance due to $M3$ in Fig. \ref{fig3}a. 
The final voltage of the two charge-sharing capacitors, $Cmain$, and $Cnew$, is the same, and the change in the voltage ($\Delta$$V$ on $Cmain$) is calculated by equation \ref{eq:2}.
\begin{equation} \label{eq:2}
\Delta V=\frac{Cunit \times (VDD - V_{0})}{Cunit + Cmain+Cnew}
\end{equation}
Due to equal charge sharing between $M3$ $(Cnew)$ and $Cmain$, the circuit does not trigger when we apply the proposed HT testing technique (earlier described in Section 5 of this paper).

\begin{figure}[!ht]
    \centering
    \includegraphics[width=0.85\linewidth]{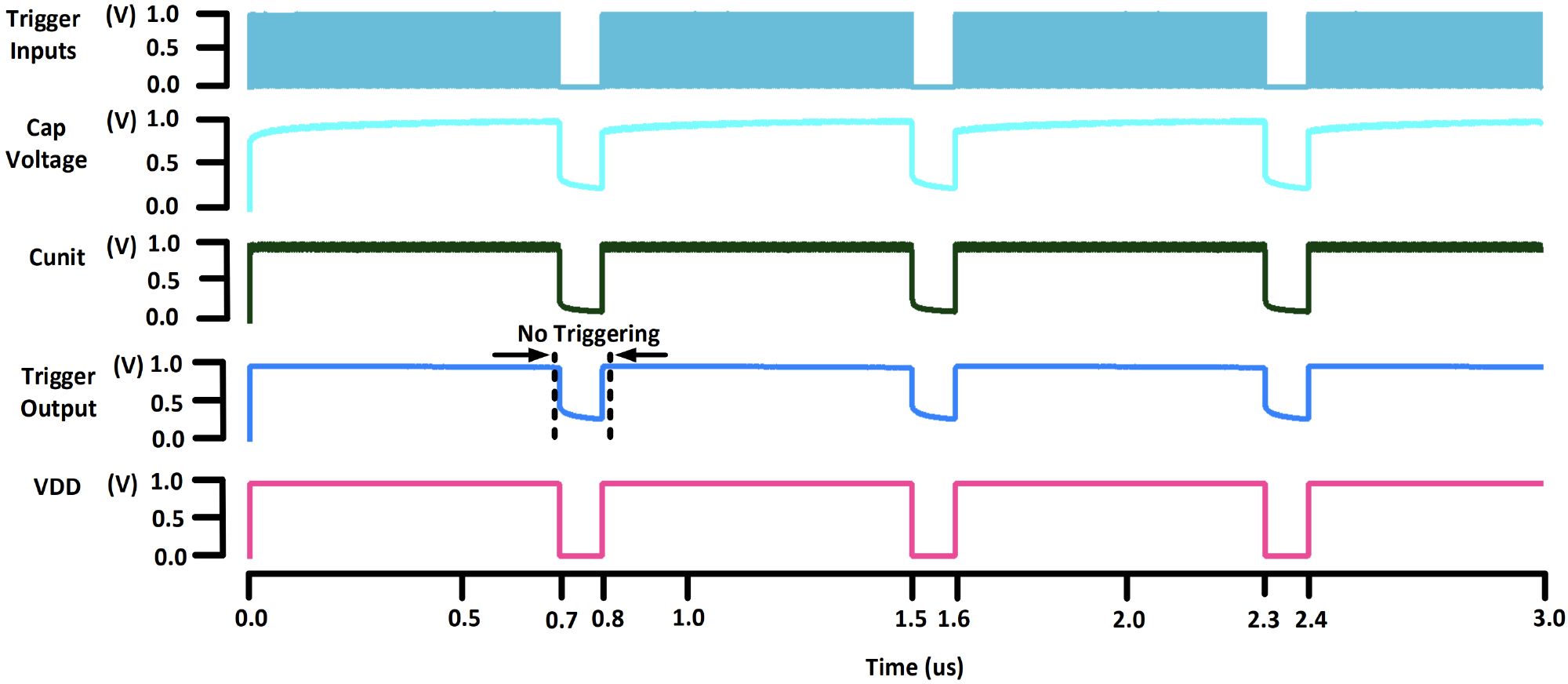}
    \caption{The timing waveform for Test Case 1 after enhancing it using the proposed AHT.}
    \label{fig11}
\end{figure}

Fig. \ref{fig11} illustrates the output waveform when the fortified Trojan circuit is tested with test case $1$.
As observed from the waveform of Fig. \ref{fig11}, the output which previously triggered now responds according to $VDD$.
This is illustrated by the two black arrows in the Trigger Output waveform in Fig. \ref{fig11}.
This fortified AHT is even more robust against our proposed $VDD$ deprivation technique. 
In order to resolve such robust AHT's, we present a novel two-phase framework to detect and mitigate our fortified AHT attacks in the following sections.

\section{Two-Phase Framework for Detecting and Mitigating AHT }\label{sec7}
\begin{figure}[ht]
    \centering
    \includegraphics[width=0.6\linewidth]{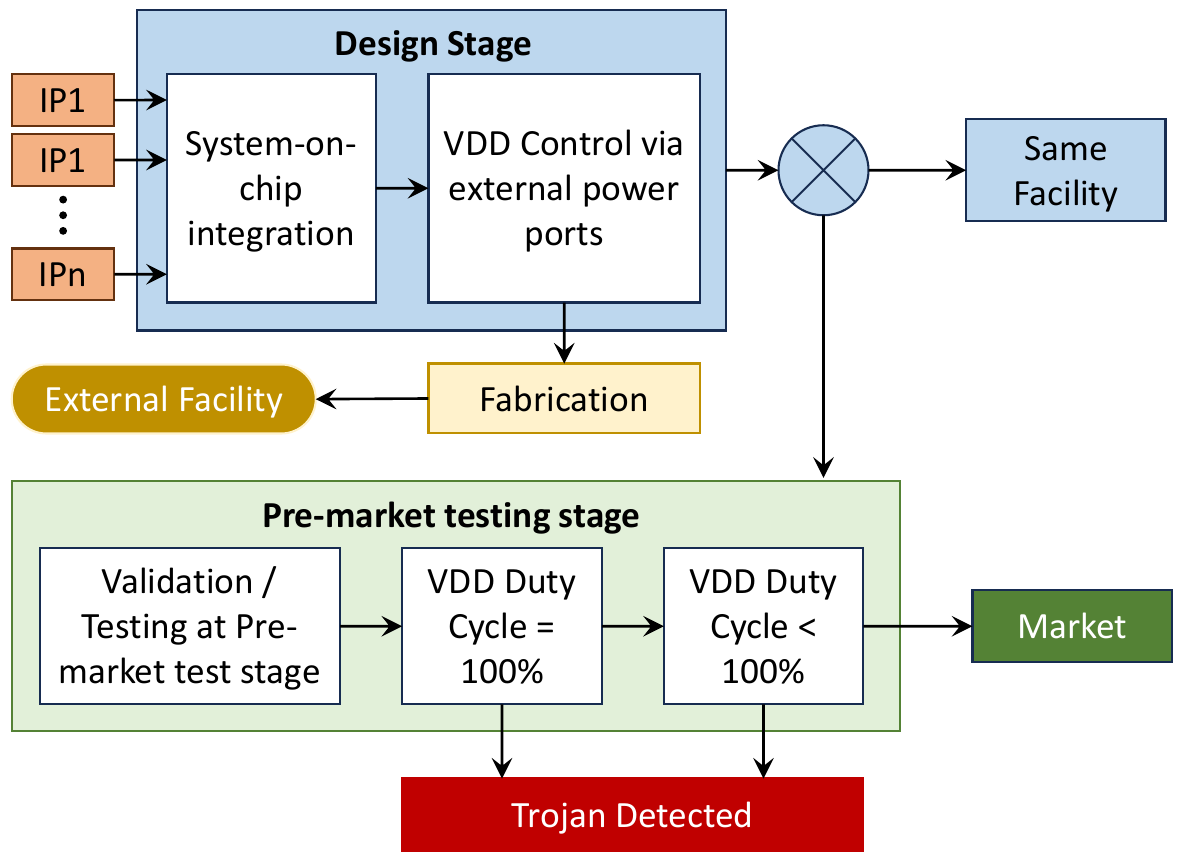}
    \caption{Our proposed framework consists of two phases: design stage (VDD control via external power ports) and pre-market testing stage (checking the VDD duty cycle).}
    \label{fig16}
\end{figure}

This section presents our proposed two-phase framework for detecting and mitigating robust AHTs. 
Fig. \ref{fig16} shows the framework which consists of the $VDD$ disruption technique ($VDD$ control) during the design phase via external power ports and pre-market test calibration during the pre-market test phase.
The design and pre-market test stages are highlighted with blue and yellow colors, respectively in Fig. \ref{fig16}.
In summary, DeMisT is composed of two phases: design-time mitigation and Pre-market validation. The first phase involves starving any possible capacitor responsible for the AHT from charging, preventing them from triggering. The second phase of our technique provides a complete framework that can detect AHT at the pre-market test stage and mitigate the effect of AHT. 

\begin{figure*}[!ht]
    \centering
    \includegraphics[width=1\linewidth]{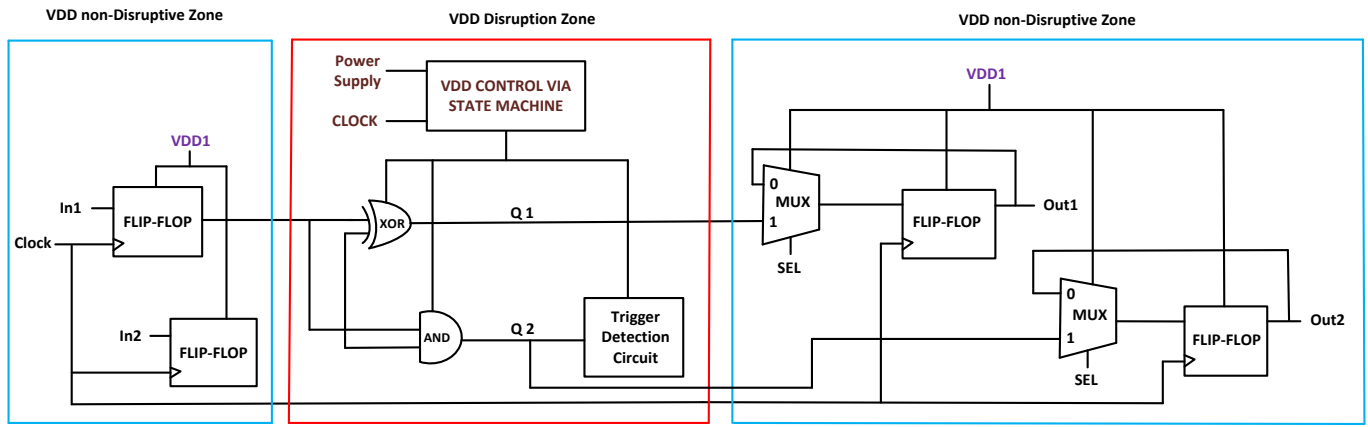}
    \caption{$VDD$ disruption in a synchronous system where PDN controls the trigger circuit}
    \label{fig13}
\end{figure*}
\begin{figure*}[!ht]
    \centering
    \includegraphics[width=0.85\linewidth]{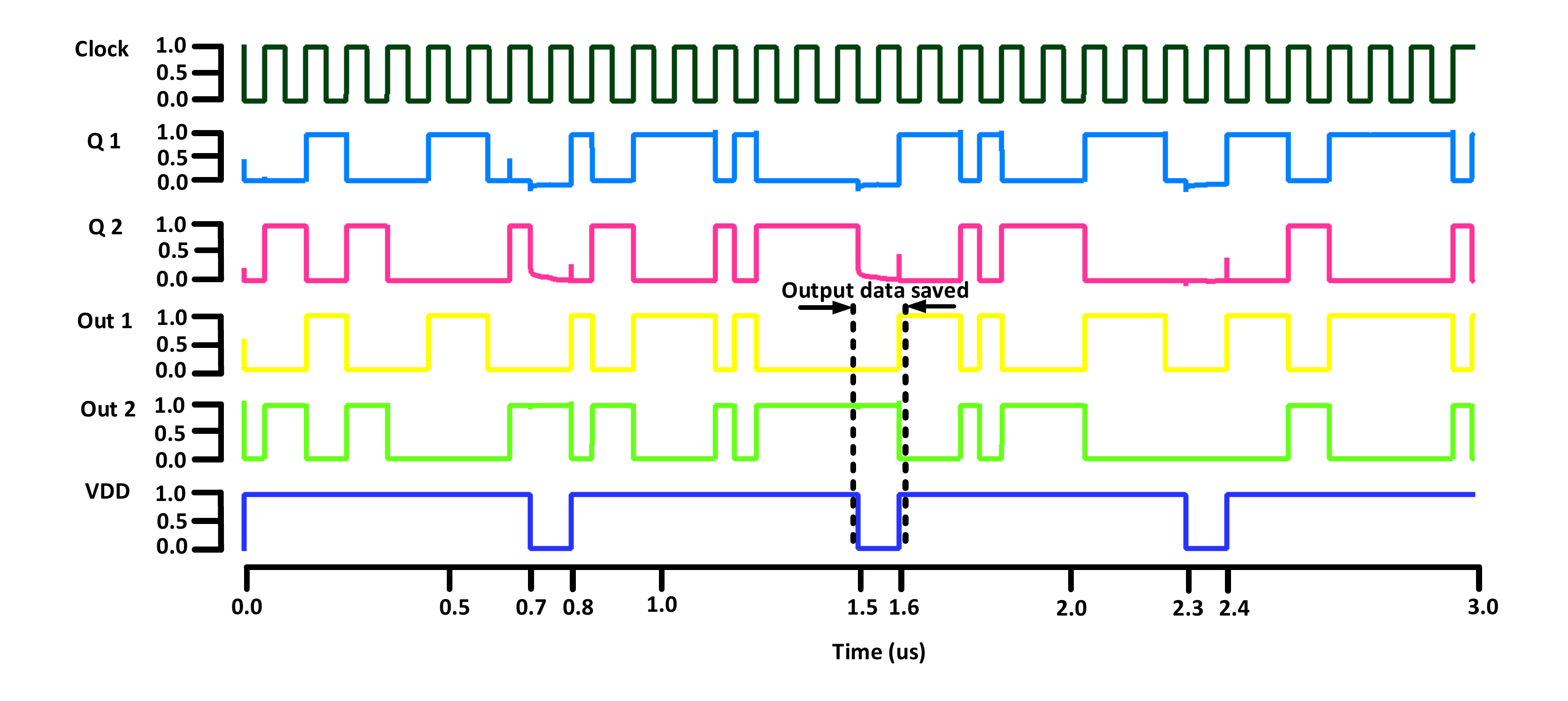}
        \caption{the experimental analysis shows that during the supply voltage ($VDD$) disruption in the intermediate combinational circuit, the data is successfully saved at the output, as shown by the black arrows (output data saved) in this figure.}
    \label{fig14}
\end{figure*}

\subsection{Design Phase}
In the design phase, the IC is activated and deployed in a sequential design where the power distribution network (PDN) is altered (i.e. a provision is provided to turn $VDD$ off for the intermediate combinational logic).
The power distribution is disrupted to prevent the capacitors responsible for the AHT from charging, hence, stopping them from triggering. In order to ensure the correct operation of the synchronous design the output data needs to be saved during the period when $VDD$ is turned off so that the synchronous system can safely revert back to its original state. We designed and implemented an efficient means of saving the data during this period as shown in Fig.~\ref{fig13}.
For proof of concept, we are showing a complete synchronous system, which can be one pipeline stage in any pipelined architecture. We divided the synchronous system into two zones based on the impact of turning off $VDD$, the $VDD$ disruption zone and the $VDD$ non-disruptive zone. 

For illustration purposes, we designed a half-adder circuit as a combinational circuit, which acts as the AHT triggering circuit in Fig. \ref{fig13}.
Our sequential circuit, i.e., $VDD$ non-disruptive zone contains a combination of D-Flip Flop (DFF) and multiplexers to maintain proper synchronous operation.
The multiplexer is used to capture the data whenever $VDD$  in the $VDD$-disruption zone is turned off.
If $SEL$ is $1$, new data is stored in the output flip-flop. 
If $SEL$ is $0$, the flip-flop stores the previous data until there is a change in $SEL$.
This ensures the output data during the time when $VDD$ is turned off in the $VDD$ Disruptive zone is saved in the $VDD$ non-disruptive zone. 
The overall implementation of the circuit is shown in Fig. \ref{fig13}.

To validate its functionality, we simulated the circuit shown in Fig.~\ref{fig13}, and the simulation waveform is illustrated in  Fig.~\ref{fig14}.
As observed from Fig.~\ref{fig14}, when $VDD$ is turned off at $700ns$, $1.5$ $\mu$$s$ and $2.3$ $\mu$$s$,  the output remains the same (it keeps the previous state).
This is indicated by the two black arrows in Fig.~\ref{fig14}.
Immediately after $VDD$ is restored, the system continues from its previous state.

\subsection {Pre-Market Test Phase}
\begin{figure}[!ht]
	\centering
	\subfloat[Algorithm for Pre-market test phase]{{\includegraphics[width=0.6\linewidth]{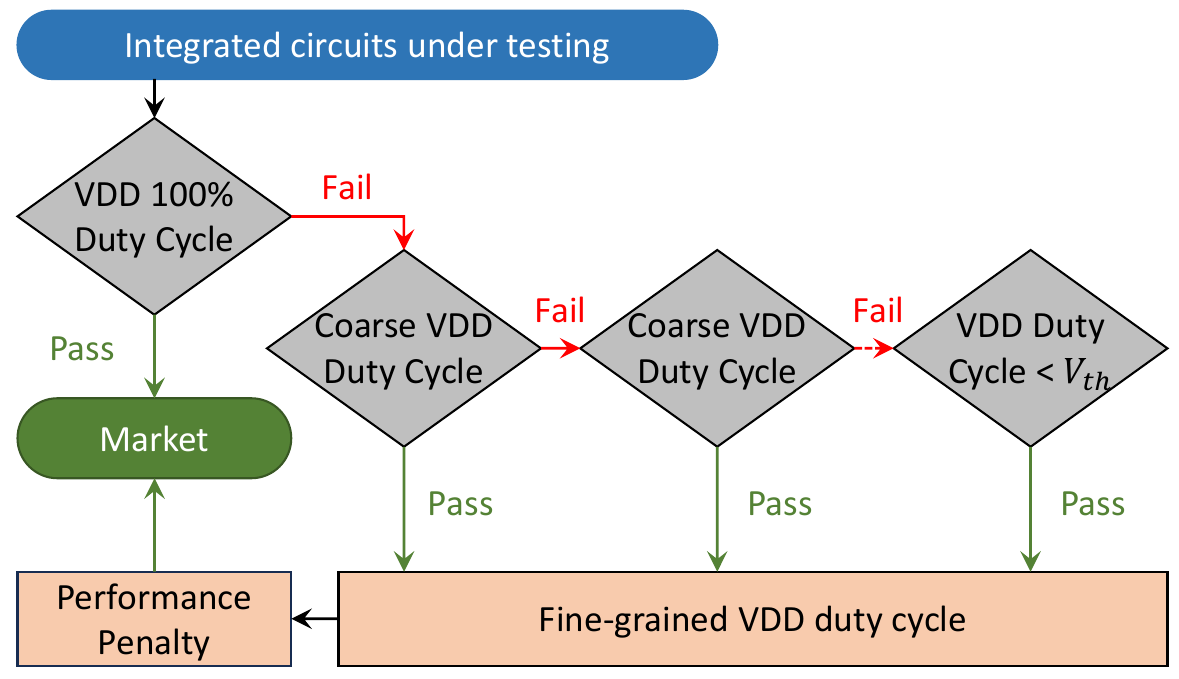} }}
	\qquad
	\subfloat[$VDD$ control mechanism]{{\includegraphics[width=0.6\linewidth]{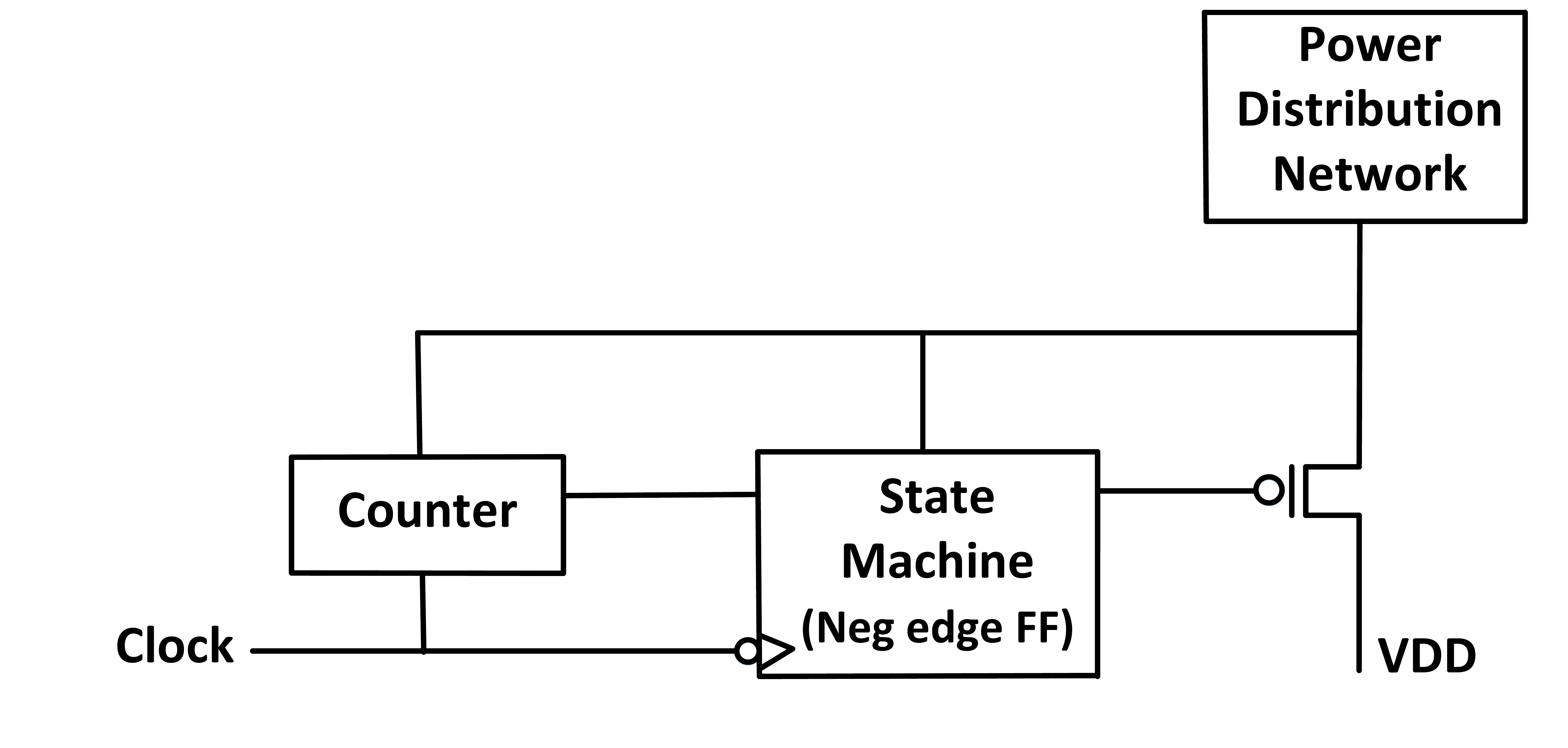} }}
	\caption{Algorithm for Pre-market test phase and $VDD$ control mechanism}
	\label{fig17}
\end{figure}

The second phase of the framework is the pre-market test phase which involves testing/validating the functionality of the IC after fabrication. This is done before the IC is released to the market.
The IC is first tested to validate the functionality with $100$\% $VDD$ and the corresponding input stimulus.
Success is determined by analyzing the power profile after the IC is excited with the input stimulus.
Ideally, the presence of the hardware Trojan leads to a change (increase) in the power profile.
Therefore, a successful test should show no change in the power profile.
Our solution can be combined with any traditional hardware Trojan detection technique that analyzes the power profile \cite{af}. For detecting AHT, our solution facilitates the triggering of such Trojans. Hence, if a failure is recorded at $100$\% $VDD$ duty cycle, further testing is done with careful calibration of $VDD$  duty cycle as shown in Fig. \ref{fig17}a. 

Fig. \ref{fig17}a shows an algorithm for the pre-market testing and hardware Trojan detection protocol, which shows that the IC is first tested with a $100$\% $VDD$ (full duty cycle) to determine whether IC is free from Trojan or not. If Trojan is found then IC is tested further with various (coarse) $VDD$ duty cycles. These duty cycles are used to determine the appropriate ratio for turning on/off time of $VDD$ to make the IC usable even if the system contains AHT. When the testing shows success for a particular $VDD$ duty cycle, the IC undergoes further testing with a fine-grained $VDD$ duty cycle. If a failure is recorded while continuously decreasing the $VDD$ duty cycle, then no further testing is performed after reaching a threshold value of the duty cycle, beyond which the performance penalty reaches unacceptable diminishing returns.

Precise $VDD$ duty cycle reduces the step size of the duty cycle, to avoid major performance penalties. 
In our experiments, it is observed that the $5$\% reduction step is a safe margin. However, a separate study needs to be done to investigate the precision of the $VDD$ duty cycle against capacitance values. 
Our proposed solution allows even the fortified AHT  to coexist with the existing synchronous system.

\subsection{Overhead Analysis}

To quantify the overhead incurred by our design, we proposed a general circuit model that controls $VDD$ distribution as shown in Fig. \ref{fig17}b.
The control circuit in Fig. \ref{fig17}, is made up of a counter, a negative edge triggered flip-flop and a PMOS transistor.
The counter can be any counter circuit that counts either increasingly or decreasingly to a maximum number.
Once this maximum number is reached, the state machine is immediately activated to switch from its current state to the next state. Many possible optimization techniques can be applied to reduce the number of bits for the counter, however, this is not the core subject of this paper, hence not discussed further in this paper. 
The state machine has two states which are implemented using a negative edge flip-flop. The negative edge ensures that the state transition occurs at the dormant stage of the clock cycle. One state sends a high signal while the other sends a low signal to the transistor to generate a specific duty cycle for $VDD$ based on the counter input. The transistor acts as a switch and the output of the state machine controls the gate of the pMOS transistor. The transistor either allows or stops $VDD$ flow from the power distribution network.

\begin{table}[!ht]
    \caption{Area overhead and number of gates for the benchmark circuits
     used to evaluate our solution}
    \begin{center}
    \resizebox{0.75\linewidth}{!}{
        \begin{tabular}{|c|c|c|c|}
        \hline
        
        \textbf{Benchmark Circuit}&\textbf{Area ($n$$m^2$)}&\textbf{Area Overhead (\%)}& \textbf{Number of Gates} \\
        
        \hline
        
        s298 & $0.079$& $63.4$ &75\\
        
        \hline
        s344 & $0.098$& $51.6$ &101\\
        \hline
        
        s349 & $0.0768$& $17.72$ &104\\
        
        \hline
        c432 & $0.158$& $3.21$ &160\\
        \hline
        
        c880 & $0.497$& $1.65$ &383\\
        \hline
        
        \end{tabular}}
    \label{tab3}
    \end{center}
\end{table}

The area overhead is determined by computing the overall circuit components of the $VDD$ non-disruptive zone in Fig. \ref{fig13}. The total components altogether are $2$ multiplexers and $7$ flip-flops. In order to ensure a fair comparison, we normalized the design using NAND gates and DFF. The area for NAND gate is estimated as ($40$$\lambda$ $\times$ $32$$\lambda$) and for a DFF as ($80$$\lambda$ $\times$ $48$$\lambda$)\cite{a27}. Table \ref{tab3} shows an evaluation of the area, number of gates and percentage of area overhead of our circuit with the various ISCAS benchmark circuits. Using ISCAS benchmark circuits \cite{a26}, we found the area overhead (second column in Table \ref{tab3}) based on the number of gates. The ratio of the overhead of our proposed solution to the overall area of the ISCAS benchmark circuit is computed in percentage and shown in column three. The fourth column shows the total number of logic gates used in these benchmark circuits. As shown in Table \ref{tab3}, the percentage of the area overhead decreases with an increase in the complexity of the micro-architecture. Hence, for a reasonable size of synchronous system (e.g. c432, or c880 are considered as one pipeline stage) our proposed solution consumes less than $4$\% of area overhead. This can be further improved with design optimization and more complex synchronous systems.

\section {Comparison with State-of-the-art HT detection techniques}
Table~\ref{tab2} shows the comparison of the proposed technique with existing state-of-the-art solutions. 
This comparative analysis is based on four parameters. Namely, the detection technique (first column), salient parameters used to detect each technique (second column), and their pros and cons (third and fourth column, respectively). The proposed technique, DeMiST, has the following advantages compared with the other state-of-the-art techniques:

\begin{enumerate}
    \item To the best of our knowledge, the proposed technique is the first one that uses the charge-depriving technique to detect such AHT that operates based on charge accumulation in a capacitor by controlling the power distribution network.
    \item Unlike Power supply signal analysis~\cite{a24} and IC fingerprinting~\cite{a25}, our proposed DeMiST technique does not require a reference Golden Trojan free model to compare the power profiles. It only relies on the increased switching activity of the triggering signal that generates abnormal power spikes. 
    \item Our technique does not require inverting the supply voltage to trigger the HT as compared to VITAMIN~\cite{a23}, rather it carefully depletes the capacitors responsible for the AHT. This charge depletion in the capacitor prevents it from triggering. 
    \item Our solution does not require keeping the circuit activity low to increase AHT sensing sensitivity. Instead, it increases the triggering frequency to such a degree that it generates huge power spikes. 
\end{enumerate}

\begin{table*}[h]
    \caption{Qualitative comparison with state-of-the-art techniques. The first column summarizes the respective technique, the second column provides the key attributes used in the detection techniques, the third column identifies the pros and finally, the last column summarizes the cons of the respective techniques.}
    \label{tab2}
    \centering
    \resizebox{1\textwidth}{!}{
        \begin{tabular}{|l|l|l|l|}
        \hline
        \rowcolor[HTML]{000000} 
        {\color[HTML]{FFFFFF} \textbf{Detection and Mitigation Techniques}} & {\color[HTML]{FFFFFF} \textbf{Key Attributes}} & {\color[HTML]{FFFFFF} \textbf{Pros.}} & {\color[HTML]{FFFFFF} \textbf{Cons.}} \\ \hline \hline
        
        \begin{tabular}[c]{@{}l@{}}\textbf{VITAMIN}~\cite{a23}: \\Inverts the supply voltage of alternate logic \\levels. Inverted voltage activates HTs with \\a higher triggering frequency. Then, by \\comparing the  power profiles, HT is detected.\end{tabular} & \begin{tabular}[c]{@{}l@{}}Sustained Vector Technique, \\ Power Profile Analysis\end{tabular} & Faster detection time & \begin{tabular}[c]{@{}l@{}}Switching between the power \\ supply voltage for each gate \\ is challenging, and a modified \\ power distribution network \\ design is required to support \\ inverted voltage.\end{tabular} \\ \hline 
        
        \begin{tabular}[c]{@{}l@{}}\textbf{Power Supply Signal Analysis}~\cite{a24}:\\The current integration technique is used to \\ analyze the IC supply current using multiple \\ supply ports.\end{tabular} & \begin{tabular}[c]{@{}l@{}}Power Supply Transient Signal,\\ Statistical Analysis\end{tabular} & Appropriate for tiny HTs & A golden IC is required. \\ \hline
        
        \begin{tabular}[c]{@{}l@{}}\textbf{IC Fingerprinting}~\cite{a25}:\\It exploits the fact that  infected circuits \\ consume more power. Hence, this technique \\ performs testing with random vectors on the \\ genuine and HT-affected ICs and compares \\ power profiles.\end{tabular} & \begin{tabular}[c]{@{}l@{}}Side-channel Analysis, \\ Signal Analysis, \\ Power Profile Analysis\end{tabular} & \begin{tabular}[c]{@{}l@{}}Higher sensitivity for \\ smaller HTs\end{tabular} & A golden IC is required. \\ \hline

        \rowcolor[HTML]{DAE8FC}
        \begin{tabular}[c]{@{}l@{}}\textbf{DeMiST} (Proposed Solution): \\It disrupts the voltage supply in the power \\ distribution network to mitigate the AHT \\ effects, and it uses the duty cycle of the supply \\ voltage to detect capacitor-based AHT.\end{tabular} & \begin{tabular}[c]{@{}l@{}}Charge Accumulation, \\ Power Distribution Network\end{tabular} & \begin{tabular}[c]{@{}l@{}}The only known solution \\ to mitigate and detect \\ capacitor-based AHT\end{tabular} & \begin{tabular}[c]{@{}l@{}}It requires access to the power \\ distribution network. \end{tabular} \\ \hline
        
        \end{tabular}
    }
\end{table*}
\section{Conclusion}\label{sec8}
In this paper, first, we proposed a charge-depriving-based AHT detection solution to detect AHTs that operate based on charge accumulation in a capacitor. Based on the experimental analysis of the proposed solution, we investigated and demonstrated the limitation of the existing state-of-the-art AHT detection circuit, i.e., $A2$. By addressing these limitations, we presented a fortified AHT that can evade the proposed AHT detection technique by reducing the required amount of charge accumulation (required to generate the triggering signals). The effectiveness of the fortified circuit has been demonstrated using simulation, corroborating the ability to evade detection. Finally, by critically analyzing the proposed fortified AHT and existing AHTs, we developed a robust two-phase framework (DeMiST) in which a synchronous system can mitigate the effects of capacitance-based stealthy AHTs by disabling the triggering capability of AHT.
Furthermore, DeMiST amplified the switching activity for charge accumulation to such a degree that it can be easily detectable using existing switching activity-based HT detection techniques. To illustrate the effectiveness, we evaluated the proposed fortified AHT and two-phase AHT detection on state-of-the-art benchmark circuits. The experimental results show that with less than 4\% area overhead, the proposed detection technique successfully detects capacitance-based AHTs. For example, the area overhead of our proposed $VDD$ control circuit is as little as $1.65$\% if introduced in a moderately complex digital logic such as ISCAS benchmark circuit c880.

%=================================================================================
\bibliographystyle{ACM-Reference-Format}

\end{document}